\documentclass[aps,prl,showpacs,showkeys,amsmath,amssymb,
twocolumn,
floatfix,
superscriptaddress
]{revtex4-1}
\usepackage{graphicx}
\usepackage{times}
\usepackage{verbatim}
\usepackage{color}
\usepackage{bm}
\usepackage{natbib}
\usepackage[normalem]{ulem}
\definecolor{gray}{rgb}{0.05,0.05,0.55}
\usepackage[colorlinks,linkcolor=gray,anchorcolor=gray,citecolor=gray, urlcolor=gray,filecolor=gray]{hyperref}

\graphicspath{{figures/}}
\begin{document}

\title{Extraction of the $^{235}$U and $^{239}$Pu Antineutrino Spectra at Daya Bay}

\newcommand{\ECUST}{\affiliation{Institute of Modern Physics, East China University of Science and Technology, Shanghai}}
\newcommand{\IHEP}{\affiliation{Institute~of~High~Energy~Physics, Beijing}}
\newcommand{\Wisconsin}{\affiliation{University~of~Wisconsin, Madison, Wisconsin 53706}}
\newcommand{\Yale}{\affiliation{Wright~Laboratory and Department~of~Physics, Yale~University, New~Haven, Connecticut 06520}} 
\newcommand{\BNL}{\affiliation{Brookhaven~National~Laboratory, Upton, New York 11973}}
\newcommand{\NTU}{\affiliation{Department of Physics, National~Taiwan~University, Taipei}}
\newcommand{\NUU}{\affiliation{National~United~University, Miao-Li}}
\newcommand{\Dubna}{\affiliation{Joint~Institute~for~Nuclear~Research, Dubna, Moscow~Region}}
\newcommand{\CalTech}{\affiliation{California~Institute~of~Technology, Pasadena, California 91125}}
\newcommand{\CUHK}{\affiliation{Chinese~University~of~Hong~Kong, Hong~Kong}}
\newcommand{\NCTU}{\affiliation{Institute~of~Physics, National~Chiao-Tung~University, Hsinchu}}
\newcommand{\NJU}{\affiliation{Nanjing~University, Nanjing}}
\newcommand{\TsingHua}{\affiliation{Department~of~Engineering~Physics, Tsinghua~University, Beijing}}
\newcommand{\SZU}{\affiliation{Shenzhen~University, Shenzhen}}
\newcommand{\NCEPU}{\affiliation{North~China~Electric~Power~University, Beijing}}
\newcommand{\Siena}{\affiliation{Siena~College, Loudonville, New York  12211}}
\newcommand{\IIT}{\affiliation{Department of Physics, Illinois~Institute~of~Technology, Chicago, Illinois  60616}}
\newcommand{\LBNL}{\affiliation{Lawrence~Berkeley~National~Laboratory, Berkeley, California 94720}}
\newcommand{\UIUC}{\affiliation{Department of Physics, University~of~Illinois~at~Urbana-Champaign, Urbana, Illinois 61801}}
\newcommand{\SJTU}{\affiliation{Department of Physics and Astronomy, Shanghai Jiao Tong University, Shanghai Laboratory for Particle Physics and Cosmology, Shanghai}}
\newcommand{\BNU}{\affiliation{Beijing~Normal~University, Beijing}}
\newcommand{\WM}{\affiliation{College~of~William~and~Mary, Williamsburg, Virginia  23187}}
\newcommand{\Princeton}{\affiliation{Joseph Henry Laboratories, Princeton~University, Princeton, New~Jersey 08544}}
\newcommand{\VirginiaTech}{\affiliation{Center for Neutrino Physics, Virginia~Tech, Blacksburg, Virginia  24061}}
\newcommand{\CIAE}{\affiliation{China~Institute~of~Atomic~Energy, Beijing}}
\newcommand{\SDU}{\affiliation{Shandong~University, Jinan}}
\newcommand{\NanKai}{\affiliation{School of Physics, Nankai~University, Tianjin}}
\newcommand{\UC}{\affiliation{Department of Physics, University~of~Cincinnati, Cincinnati, Ohio 45221}}
\newcommand{\DGUT}{\affiliation{Dongguan~University~of~Technology, Dongguan}}
\newcommand{\XJTU}{\affiliation{Department of Nuclear Science and Technology, School of Energy and Power Engineering, Xi'an Jiaotong University, Xi'an}}
\newcommand{\UCB}{\affiliation{Department of Physics, University~of~California, Berkeley, California  94720}}
\newcommand{\HKU}{\affiliation{Department of Physics, The~University~of~Hong~Kong, Pokfulam, Hong~Kong}}
\newcommand{\UH}{\affiliation{Department of Physics, University~of~Houston, Houston, Texas  77204}}
\newcommand{\Charles}{\affiliation{Charles~University, Faculty~of~Mathematics~and~Physics, Prague}} 
\newcommand{\USTC}{\affiliation{University~of~Science~and~Technology~of~China, Hefei}}
\newcommand{\TempleUniversity}{\affiliation{Department~of~Physics, College~of~Science~and~Technology, Temple~University, Philadelphia, Pennsylvania  19122}}
\newcommand{\CUC}{\affiliation{Instituto de F\'isica, Pontificia Universidad Cat\'olica de Chile, Santiago}} 
\newcommand{\CGNPG}{\affiliation{China General Nuclear Power Group, Shenzhen}}
\newcommand{\NUDT}{\affiliation{College of Electronic Science and Engineering, National University of Defense Technology, Changsha}} 
\newcommand{\IowaState}{\affiliation{Iowa~State~University, Ames, Iowa  50011}}
\newcommand{\ZSU}{\affiliation{Sun Yat-Sen (Zhongshan) University, Guangzhou}}
\newcommand{\CQU}{\affiliation{Chongqing University, Chongqing}} 
\newcommand{\BCC}{\altaffiliation[Now at ]{Department of Chemistry and Chemical Technology, Bronx Community College, Bronx, New York  10453}} 

\newcommand{\UCI}{\affiliation{Department of Physics and Astronomy, University of California, Irvine, California 92697}} 
\author{D.~Adey}\IHEP
\author{F.~P.~An}\ECUST
\author{A.~B.~Balantekin}\Wisconsin
\author{H.~R.~Band}\Yale
\author{M.~Bishai}\BNL
\author{S.~Blyth}\NTU
\author{D.~Cao}\NJU
\author{G.~F.~Cao}\IHEP
\author{J.~Cao}\IHEP
\author{J.~F.~Chang}\IHEP
% ...

\collaboration{The Daya Bay Collaboration}\noaffiliation
\date{\today}

\begin{abstract}
This Letter reports the first extraction of individual antineutrino spectra from $^{235}$U and $^{239}$Pu fission and an improved measurement of the prompt energy spectrum of reactor antineutrinos at Daya Bay.
The analysis uses $3.5\times 10^6$ inverse beta-decay candidates in four near antineutrino detectors in 1958 days.
The individual antineutrino spectra of the two dominant isotopes, $^{235}$U and $^{239}$Pu, are extracted using the evolution of the prompt spectrum as a function of the isotope fission fractions.
In the energy window of 4--6~MeV, a 7\% (9\%) excess of events is observed for the $^{235}$U ($^{239}$Pu) spectrum compared with the normalized Huber-Mueller model prediction.
The significance of discrepancy is $4.0\sigma$ for $^{235}$U spectral shape compared with the Huber-Mueller model prediction.
The shape of the measured inverse beta-decay prompt energy spectrum disagrees with the prediction of the Huber-Mueller model at $5.3\sigma$.
In the energy range of 4--6~MeV, a maximal local discrepancy of $6.3\sigma$ is observed.
\end{abstract}

\maketitle
Nuclear reactors are powerful sources of electron antineutrinos ($\bar\nu_e$) and have played an important role in neutrino physics.
Most recently, Daya Bay~\cite{bib:prl_rate,bib:cpc_osc,bib:prl2014, bib:prl2015, bib:prd1230,bib:prl1958}, RENO~\cite{bib:RENO,bib:reno2018}, and Double Chooz~\cite{bib:DoubleChooz,bib:DoubleChooz2018} Collaborations reported observations of neutrino oscillation induced by a nonzero mixing angle $\theta_{13}$.
In addition, these experiments also provided measurements of reactor $\bar\nu_e$ flux and spectrum~\cite{bib:prl_reactor, bib:cpc_reactor, bib:reno_bump,bib:dc_bump} at distances of 300--500~m from the reactors.
The flux measurements confirmed the $\sim$6\% deficit found in the 2011 reevaluation~\cite{bib:mueller2011, bib:huber} of the reactor $\bar\nu_e$ flux (``reactor antineutrino anomaly''~\cite{bib:mention2011}).
The spectral measurements indicated a new anomaly (``5-MeV bump'') when compared with theoretical calculations, an observation further confirmed by the NEOS Collaboration~\cite{bib:NEOS}, and by reexamination of earlier reactor antineutrino data~\cite{bib:Goesgen}.
Observation of the evolution of the reactor $\bar\nu_e$ spectrum from commercial reactors~\cite{bib:old,bib:evolution,bib:RENOevolution,bib:DANSS} and measurement of the $^{235}$U $\bar\nu_e$ spectrum from highly enriched uranium research reactors~\cite{bib:ILLspectrum,bib:PROSPECT} have also been performed, providing first glimpses at the dependence of spectral features on reactor fuel content.
Interpretations of the reactor $\bar\nu_e$ flux and spectrum anomalies reveal the complexes in the fission
beta spectrum conversion and nuclear databases~\cite{Hayes:2013wra, Dwyer:2014eka, Fang:2015cma, Hayes:2015yka, Sonzogni:2016yac, Sonzogni:2017wxy, Gebre:2017vmm}.
Additional precision measurements are essential to fully investigate the origins of the reactor $\bar\nu_e$ flux and spectrum anomalies, and provide crucial inputs to future reactor neutrino experiments~\cite{bib:JUNO}.

This Letter reports the extracted individual prompt energy spectra of two dominant isotopes ($^{235}$U and $^{239}$Pu) for the first time by fitting the evolution of the prompt energy spectrum as a function of fission fractions from commercial reactors.
In addition, an improved measurement of the prompt energy spectrum of reactor $\bar\nu_e$ is reported with three times more $\bar\nu_e$ events and reduced systematic uncertainties compared with previous results~\cite{bib:cpc_reactor}.

The Daya Bay Reactor Neutrino Experiment is located near the Daya Bay nuclear power plant, which hosts six commercial pressurized-water reactors (2.9~GW maximum thermal power).
Identically designed $\bar\nu_e$ detectors (ADs) are deployed in two near halls (EH1 and EH2) containing two ADs each and in the far hall (EH3) with four ADs.
The analysis uses 1958 days of data from four near ADs. 
Details about the experiment and the data set are given in Refs.~\cite{bib:detector, bib:prl1958}.

In a typical commercial reactor, antineutrinos are produced from thousands of beta-decay branches of the fission products from four major isotopes, $^{235}$U, $^{239}$Pu, $^{238}$U, and $^{241}$Pu.
The $\bar\nu_e$ spectrum is measured with inverse beta-decay (IBD) reactions: $\bar{\nu}_{e}+p\rightarrow e^+ + n$.
The predicted $\bar\nu_e$ energy spectrum in a detector at a given time $t$ is calculated as
\begin{equation}\label{equ_generic}
    S_d(E_{\nu}, t)  = N_d\epsilon_d\sigma(E_{\nu})\sum_r \frac{P_{ee}(E_{\nu},L_{rd})}{ 4\pi L^2_{rd}}\frac{d\phi_r(E_{\nu},t)}{dE_{\nu}dt},
\end{equation}
where $E_{\nu}$ is the $\bar\nu_e$ energy, $d$ is the detector index, $r$ is the reactor index, $N_d$ is the target proton number, $\epsilon_d$ is the detection efficiency, $L_{rd}$ is the distance from detector $d$ to reactor $r$, $P_{ee}(E_{\nu},L_{rd})$ is the $\bar{\nu}_e$ survival probability in the standard three-neutrino model, and $\sigma(E_{\nu})$ is the IBD cross section.
The energy spectrum of antineutrinos from one reactor is
\begin{equation}\label{equ_prediction}
\begin{split}
\frac{d\phi_r(E_{\nu},t)}{dE_{\nu}dt} &= \frac{W_r(t)}{\sum_i f_{ir}(t) e_i}\sum_i f_{ir}(t)s_i(E_{\nu})c_i^{ne}(E_{\nu},t) \\
&+ s_{\mathrm{SNF}}(E_{\nu},t) + s_{\mathrm{NL}}(E_{\nu},t),
\end{split}
\end{equation}
where $W_r(t)$ is the thermal power of reactor $r$, $e_i$ is the energy released per fission for isotope $i$, $f_{ir}(t)$ is the fission fraction, $s_i(E_{\nu})$ is the $\bar\nu_e$ energy spectrum per fission for each isotope, $c_i^{ne}(E_{\nu},t)$ is a function of the order of unity absorbing the correction due to nonequilibrium effects, $s_{\mathrm{SNF}}(E_{\nu},t)$ and $s_{\mathrm{NL}}(E_{\nu},t)$ are contributions from spent nuclear fuel (SNF) and from nuclides with $\bar\nu_e$ flux with a nonlinear dependence on reactor neutron flux~\cite{bib:neutron_nonlinear}, respectively.

For $s_i(E_{\nu})$ in Eq.~\ref{equ_prediction}, the $^{235}$U, $^{239}$Pu, and $^{241}$Pu $\bar\nu_e$ spectra from Huber~\cite{bib:huber} and $^{238}$U spectrum from Mueller~\cite{bib:mueller2011} are used in the prediction (Huber-Mueller model).
Thermal power and fission fraction data are provided by the Daya Bay nuclear power plant with uncertainties of 0.5\% and 5\%~\cite{bib:cpc_reactor}, respectively.
The correlations of fission fractions among the four isotopes are taken from Ref.~\cite{bib:cpc_reactor}.
The energies released per fission ($e_i$) are taken from Ref.~\cite{bib:maxb_fissionenergy}.

In contrast to previous Daya Bay analyses, the nonequilibrium correction and contributions from SNF and nonlinear nuclides are estimated and added to the flux prediction with time evolution.
The nonequilibrium effect exists for ILL measurements~\cite{bib:ILL_1,bib:ILL_2,bib:ILL_3}, which are the basis of the Huber-Mueller model, due to a limited irradiation time.
The correction of the nonequilibrium effect ($0.7\%$) for each batch of fuel elements is calculated daily based on the irradiation time~\cite{bib:mueller2011}.
The SNF ($0.2\%$), including contribution from the storage water pool and the shutdown reactor core, is calculated daily using the refueling history provided by the power plant.
The $\bar\nu_e$ flux from some nuclides has a nonlinear dependence on the neutron flux in a reactor core~\cite{bib:neutron_nonlinear}.
The correction for these nonlinear nuclides is obtained as a function of time and contributes $<\!0.1\%$ of the total $\bar\nu_e$ flux.

The $3.5\times 10^6$ IBD candidates in the four near ADs and the expected backgrounds from Ref.~\cite{bib:prl1958} are used in this analysis.
The accidental and Am-C correlated backgrounds are estimated daily in each AD.
The cosmogenic $^9$Li/$^8$He, fast neutron, and $^{13}$C($\alpha$, n)$^{16}$O backgrounds are treated as constants in time.
The IBD detection efficiency is 80.25\% with a correlated uncertainty of 1.19\%~\cite{bib:flux_eff_paper} and an uncorrelated uncertainty of 0.13\% among ADs.
The oscillation parameters $\sin^22\theta_{13}=0.0856\pm 0.0029$ and $\Delta m^2_{ee} = (2.522^{+0.068}_{-0.070})\times 10^{-3}$~eV$^2$ from Ref.~\cite{bib:prl1958} are used to correct for the oscillation effect, namely $P_{ee}(E_{\nu},L_{rd})$ in Eq.~\ref{equ_generic}.

The predicted prompt energy spectrum is determined from the $\bar\nu_e$ spectrum taking into account the effects of IBD kinematics, energy leakage, and energy resolution.
A model of the nonlinear energy response is used to correct the measured prompt energy spectrum of the IBD candidates~\cite{bib:nonlinear} to facilitate the comparison of spectra between different experiments~\footnote{In the previous spectrum measurements~\cite{bib:prl_reactor,bib:cpc_reactor}, the energy response was not corrected in the measured prompt energy spectrum, but it was applied to the predicted energy spectrum.}.
The magnitude of the nonlinear correction is $\sim$10\% at maximum with a 0.5\% uncertainty at 3 MeV~\cite{bib:nonlinear}, improved from 1\% previously~\cite{bib:cpc_reactor}.

\begin{figure}[htbp]
\centering
\includegraphics[width=\columnwidth]{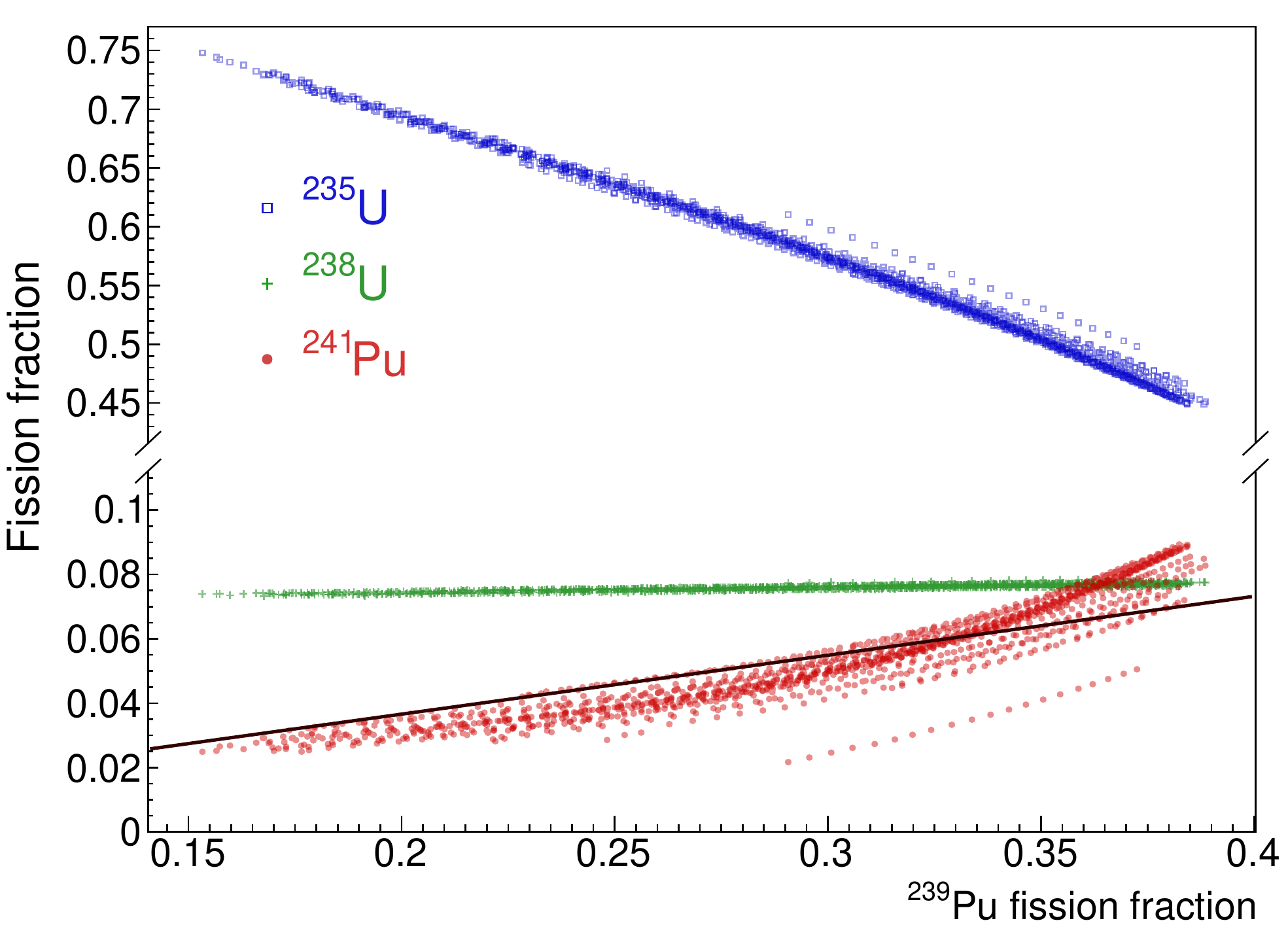}
\caption{The weekly fission fractions for the four major isotopes in the six reactors in 1958 days including four to six refueling cycles for each. The solid line represents an approximately linear relation between fission fractions of $^{239}$Pu and $^{241}$Pu.}
\label{fission_fraction}
\end{figure}

The evolution of fission fractions of the four major isotopes in multiple refueling cycles is shown in Fig.~\ref{fission_fraction} for the six reactors during operation.
The dominant isotopes contributing to the prompt spectrum are $^{235}$U and $^{239}$Pu, as their fission fractions add up to $\sim$87\%.

Each isotope produces a unique $\bar\nu_e$ spectrum depending on its fission products and corresponding beta-decay spectra~\cite{Fallot:2012jv, bib:HayesVogel}.
Since the observed prompt energy spectrum in one AD is a combination of the individual spectra of four isotopes, it evolves as a function of fission fractions~\cite{bib:evolution, bib:RENOevolution, Hayes:2017res, bib:DYB-NEOS}.
In order to describe the relative contribution of each isotope in one AD from the six reactors, we define an effective fission fraction for isotope $i$ observed by detector $d$ as
\begin{equation}
f^{\rm eff}_{id} (t) = \sum_r \frac{W_r(t)f_{ir}(t)}{L^2_{rd}\sum_j f_{jr}(t) e_j }/\sum_r \frac{W_r(t)}{L^2_{rd}\sum_j f_{jr}(t) e_j }.
\end{equation}
The variation of detectorwise effective fission fraction of $^{235}$U ($^{239}$Pu) is 50\%--65\% (24\%--35\%), smaller than the variation of reactorwise fission fraction shown in Fig.~\ref{fission_fraction}.

The 1958 days of data are divided into 20 groups ordered by the $^{239}$Pu effective fission fraction in each week for each AD.
The evolution of the prompt energy spectrum is dominated by $^{235}$U and $^{239}$Pu, while it is less sensitive to $^{238}$U and $^{241}$Pu due to smaller fission fractions.
To extract the individual spectra of the $^{235}$U and $^{239}$Pu isotopes, $s^5(\boldsymbol{\eta}^{5})$ and $s^9(\boldsymbol{\eta}^{9})$, respectively, from the prompt energy spectrum, a $\chi^2$ function in the Poisson-distributed form is constructed as
\begin{equation}
\chi^2(\boldsymbol{\eta}^{5}, \boldsymbol{\eta}^{9}) \!= \! 2\sum_{djk}(S_{djk}\!-\! M_{djk}\!+\! M_{djk}\ln\frac{M_{djk}}{S_{djk}}) \!+\! f(\boldsymbol{\epsilon}, \mathbf{\Sigma}),
\end{equation}
where $d$ is the detector index, $j$ is the index of the data groups, $k$ is the prompt energy bin, $M_{djk}$ is the measured prompt energy spectrum of each data group, $\boldsymbol{\epsilon}$ is a set of nuisance parameters, $f(\boldsymbol{\epsilon}, \mathbf{\Sigma})$ is the term to constrain the nuisance parameters incorporating systematic uncertainties and their correlations ($\mathbf{\Sigma}$) among the reactors, detectors, and data groups, and
\begin{equation}
S_{djk} \!=\! \alpha_k(\boldsymbol{\epsilon})s^5_k(\eta_k^{5}) \!+\! \beta_k(\boldsymbol{\epsilon})s^9_k(\eta_k^{9}) \!+\! s_k^{238+241}(\boldsymbol{\epsilon}) \!+\! c_k(\boldsymbol{\epsilon})
\end{equation}
is the corresponding expected prompt energy spectrum without normalization, $s^5_k(\eta_k^{5})$ [$s^9_k(\eta_k^{9})$] is the element of extracted $^{235}$U ($^{239}$Pu) spectrum at energy bin $k$, $\alpha_k(\boldsymbol{\epsilon})$ [$\beta_k(\boldsymbol{\epsilon})$] is the corresponding coefficient for the $^{235}$U ($^{239}$Pu) taking into account the detector target mass, detection efficiency, baseline, and number of fissions,
$s_k^{238+241}(\boldsymbol{\epsilon})$ is the expected prompt energy spectra contributed from $^{238}$U and $^{241}$Pu, and $c_k(\boldsymbol{\epsilon})$ includes contributions from the SNF, nonlinear nuclides, and backgrounds.
The Huber-Mueller flux model is used to calculate the initial prompt energy spectrum for the four isotopes.
Two sets of free parameters, $\boldsymbol{\eta}^5$ and $\boldsymbol{\eta}^9$, are applied to the 26 energy bins correcting the initial $^{235}$U and $^{239}$Pu spectra, respectively.
As a result, the individual $^{235}$U and $^{239}$Pu spectra corrected with the best fit values of $\boldsymbol{\eta}^5$ and $\boldsymbol{\eta}^9$ do not depend on the input of the initial spectra.
For the $^{238}$U and $^{241}$Pu spectra, nuisance parameters are incorporated in each energy bin to vary the initial spectra within their uncertainties.
We conservatively enlarge the uncertainties of the $^{238}$U and $^{241}$Pu spectra quoted in the Huber-Mueller model based on the investigations of the antineutrino spectrum evaluations from nuclear databases~\cite{bib:mueller2011, bib:mention2011}.
For the $^{238}$U spectrum, the uncertainty is 15\% in 0.7--4.5~MeV, 20\% in 4.5--6~MeV, 30\% in 6--7~MeV, and 60\% in 7--8~MeV, and for $^{241}$Pu it is 10\% in 0.7--7~MeV and 50\% in 7--8~MeV.
Additional normalization uncertainties of 15\% and 10\%~\cite{bib:evolution} are assigned to the  $^{238}$U and $^{241}$Pu spectra, respectively.

The time dependence of reactor antineutrino production and detector response, and their impact on the $^{235}$U and $^{239}$Pu spectra, are examined.
The drift of the energy scale is controlled to $<\!0.1\%$ and the relative variation of energy resolution in the 20 data groups is 3\%.
Therefore, the detector energy response~\cite{bib:cpc_reactor} is treated as stable with its uncertainty treated as time independent.
The uncertainties of reactor power and fission fractions are treated as correlated between the data groups, and treating them as uncorrelated has a negligible effect in this analysis.

Performing the $\chi^2$ fit with one energy bin covering the whole spectrum (0.7--8~MeV), we obtain the IBD yields of $(6.10\pm 0.15)\times 10^{-43}$~cm$^2$/fission and $(4.32\pm 0.25)\times 10^{-43}$~cm$^2$/fission for $^{235}$U and $^{239}$Pu, respectively, with a $\chi^2/ndf = 88/78$.
The ratios to the expected IBD yield from the Huber-Mueller model are $0.920\pm0.023(\mathrm{exp.})\pm0.021(\mathrm{model})$ and $0.990\pm0.057(\mathrm{exp.})\pm0.025(\mathrm{model})$ for $^{235}$U and $^{239}$Pu, respectively, consistent with the previous analysis~\cite{bib:evolution}.
Removing the time dependence of the nonequilibrium effect, SNF, and nonlinear nuclides produces a shift of less than 0.7\% in the IBD yields of $^{235}$U and $^{239}$Pu.

\begin{figure}[htbp]
    \centering
    \includegraphics[width=\columnwidth]{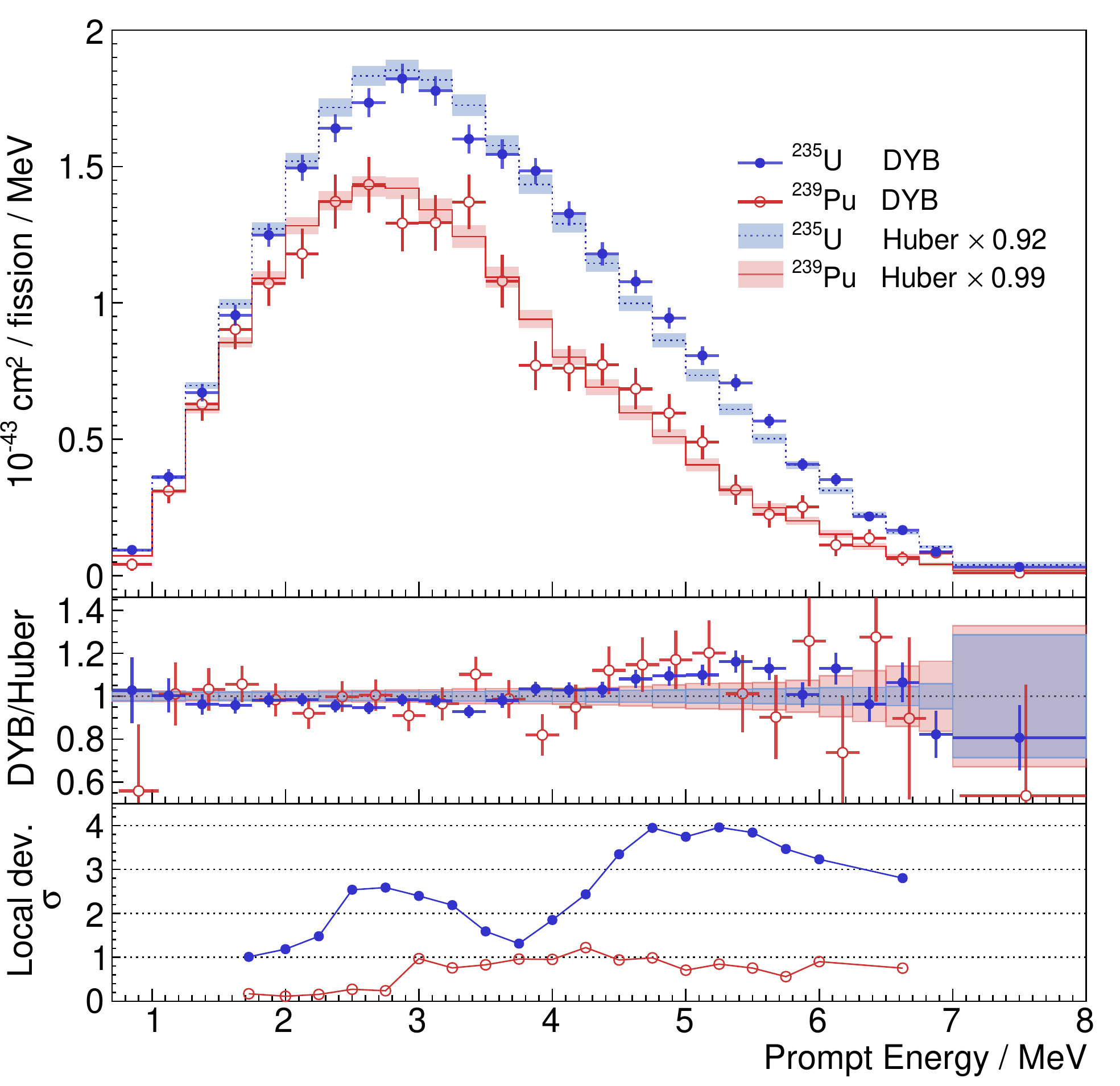}
    \caption{(Top) Comparison of the extracted $^{235}$U and $^{239}$Pu spectra and the corresponding Huber-Mueller model predictions with the normalization factors 0.92 and 0.99, respectively.
    The error bars in the data points are the square root of the diagonal terms of the covariance matrix of the extracted spectra.
    The error bands are the uncertainties from the Huber-Mueller model.
    (Middle) Ratio of the extracted spectra to the predicted spectra. The $^{239}$Pu data points are displaced for visual clarity of error bars.
    (Bottom) Local significance of the shape deviations for the extracted $^{235}$U and $^{239}$Pu spectra compared to the model predictions.}
    \label{extracted_spectrum}
\end{figure}

The top panel of Fig.~\ref{extracted_spectrum} shows the extracted $^{235}$U and $^{239}$Pu spectra together with their Huber-Mueller predictions normalized to the best-fit numbers of events for $^{235}$U (0.920) and $^{239}$Pu (0.990), respectively.
In the middle panel, the ratios of the extracted spectra to the corresponding predicted spectra are shown.
An edge around 4~MeV is found in the $^{239}$Pu spectrum compared to the prediction.
Analysis with a different data grouping, or analysis with only EH1 or EH2 data shows a similar edge.
In the energy window of 4--6~MeV, a 7\% (9\%) excess of events is observed for $^{235}$U ($^{239}$Pu) spectrum compared with the normalized Huber-Mueller model prediction.
A $\chi^2$ test is performed to quantify the local discrepancy between the extracted $^{235}$U and $^{239}$Pu spectra and their corresponding predicted spectra following the method in Ref.~\cite{bib:cpc_reactor}.
As shown in the bottom panel of Fig.~\ref{extracted_spectrum}, the features of the $^{239}$Pu in 3--4~MeV show a 1$\sigma$ local discrepancy.
The maximum local discrepancy is $4.0\sigma$ for the $^{235}$U spectrum, and only $1.2\sigma$ for the $^{239}$Pu spectrum because of larger uncertainties.
If the $^{239}$Pu spectrum is fixed to have the same spectral shape discrepancy as the $^{235}$U spectrum in 4--6~MeV, we obtain a change in the $\chi^2$ value, $\Delta \chi^2/ndf$ = 4.0/8, corresponding to a 0.2$\sigma$ inconsistency.
Thus, the Daya Bay data indicate an incorrect prediction of the $^{235}$U spectrum, but such a conclusion cannot be drawn for the other primary fission isotopes.
Combining the results of IBD yield and spectral shape, we deduce that the 8\% deficit of $^{235}$U IBD yield is dominated by the deficit in the energy range below 4~MeV with a significance of 4$\sigma$ with respect to the Huber-Mueller model prediction without normalization.

\begin{figure}[htbp]
    \centering
    \includegraphics[width=\columnwidth]{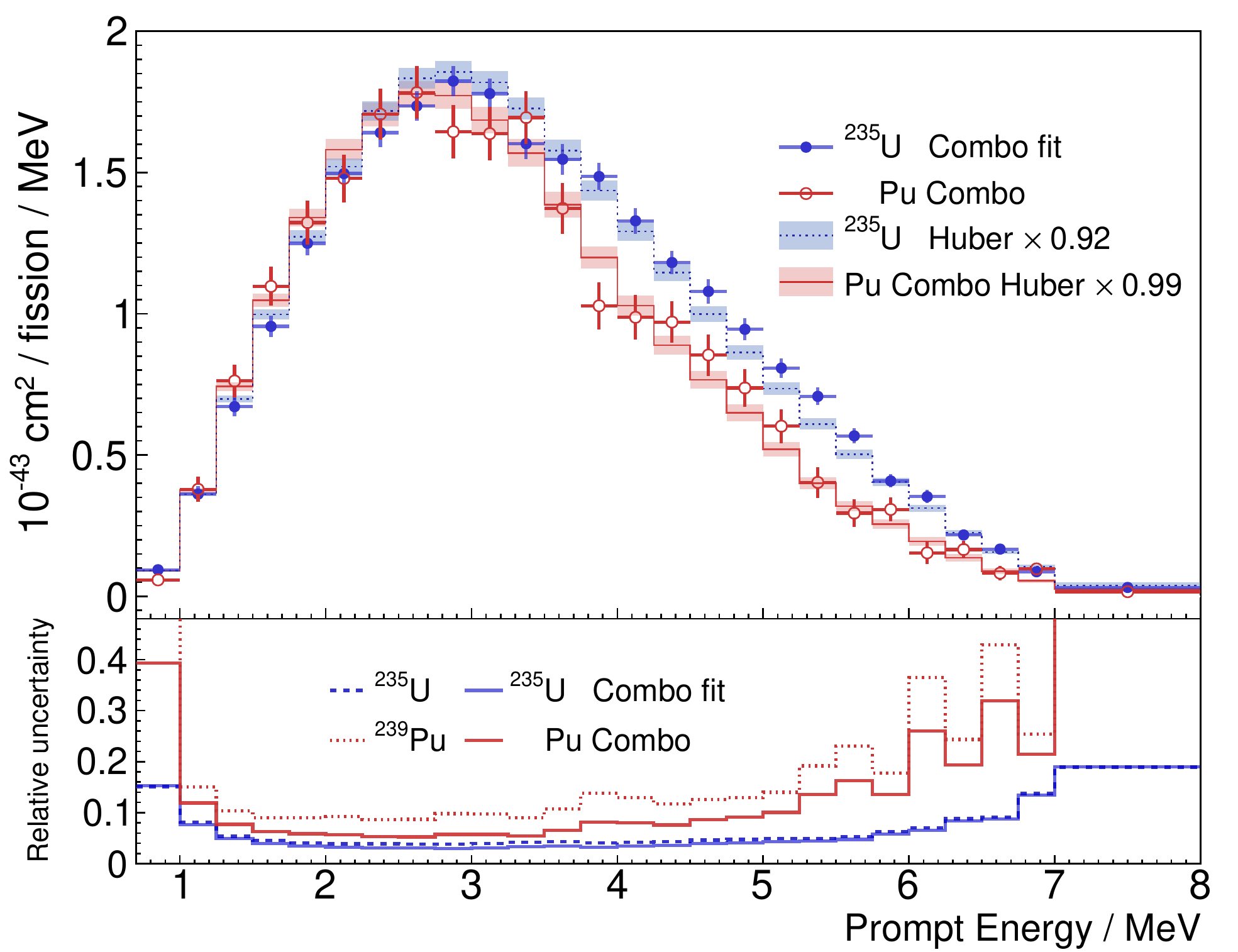}
    \caption{(Top) Comparison of the extracted $^{235}$U spectrum and $s_{\mathrm{combo}}$ as a combination of $^{239}$Pu and $^{241}$Pu with the corresponding Huber-Mueller predicted spectra with the normalization factors 0.92 and 0.99. (Bottom) The fractional size of the diagonal elements of the covariance matrix for extracted spectra with and without the combination of $^{239}$Pu and $^{241}$Pu.}
    \label{extracted_spectrum_combo}
\end{figure}

The fractional size of the diagonal elements of the covariance matrix is shown in the bottom panel of Fig.~\ref{extracted_spectrum_combo}, which is $4\%$ for $^{235}$U and $9\%$ for $^{239}$Pu around 3~MeV.
The statistical uncertainty contributes to about 55\% (60\%) of the total uncertainty of $^{235}$U ($^{239}$Pu).
The uncertainties from the input $^{238}$U and $^{241}$Pu spectra and rates contribute about 35\% for both $^{235}$U and $^{239}$Pu.
The other uncertainties contribute to about $10\%$ (5\%) for $^{235}$U ($^{239}$Pu).
The spectral uncertainties of $^{235}$U and $^{239}$Pu are anticorrelated with correlation coefficients between $-0.8$ and $-0.3$.
The $^{235}$U and $^{239}$Pu spectra as well as their associated covariance matrix are provided in the Supplemental Material~\cite{bib:Supple}.
An independent analysis based on Bayesian inference using Markov Chain Monte Carlo calculations with different data grouping obtains consistent results.

The extracted spectra of $^{235}$U and $^{239}$Pu have a certain dependence on the inputs of the $^{238}$U and $^{241}$Pu spectra.
The fission fraction of $^{241}$Pu is approximately proportional to $^{239}$Pu as shown in Fig.~\ref{fission_fraction}, thus, they can be treated as one component in the contribution to the prompt energy spectrum.
A combination of $^{239}$Pu and $^{241}$Pu spectra ($s_{239}$ and $s_{241}$), as an invariant spectrum independent of the fission fractions, is defined as $s_{\mathrm{combo}} = s_{239} + 0.183\times s_{241}$.
The coefficient of 0.183 is the average fission fraction ratio of $^{241}$Pu to $^{239}$Pu in 1958 days, shown as a line in Fig.~\ref{fission_fraction}.
The residual contribution of $^{241}$Pu spectrum is corrected using the Huber-Mueller model for some data groups when the fission fraction ratios of $^{241}$Pu to $^{239}$Pu deviate from 0.183.
With this combination of $^{239}$Pu and $^{241}$Pu, the dependence on the input $^{241}$Pu spectrum is largely removed.
The top panel of Fig.~\ref{extracted_spectrum_combo} shows the extracted $^{235}$U spectrum and $s_{\mathrm{combo}}$ compared with the normalized Huber-Mueller model predictions.
The bottom panel shows the uncertainties of extracted spectra.
The uncertainty of $s_{\mathrm{combo}}$ is 6\% around 3~MeV, improved from 9\% in the case of no combination.
The extracted $s_{\mathrm{combo}}$ can be used to predict the $\bar\nu_e$ spectrum in experiments with a similar fission fraction ratio of $^{241}$Pu to $^{239}$Pu.

The time-averaged IBD yield is measured to be $(5.94\pm 0.09)\times 10^{-43}$~cm$^2$/fission, where the statistical uncertainty is 0.05\% and the systematic uncertainty is 1.5\% taken from Table~1 in Ref.~\cite{bib:flux_eff_paper}.
The corresponding average fission fractions for the four major isotopes $^{235}$U, $^{239}$Pu, $^{238}$U, and $^{241}$Pu are 0.564, 0.304, 0.076, 0.056, respectively.
The ratio of the measured IBD yield to the Huber-Mueller model prediction is $0.953\pm 0.014$ (exp.) $\pm 0.023 $ (model).

\begin{figure}[htbp]
\centering
\includegraphics[width=\columnwidth]{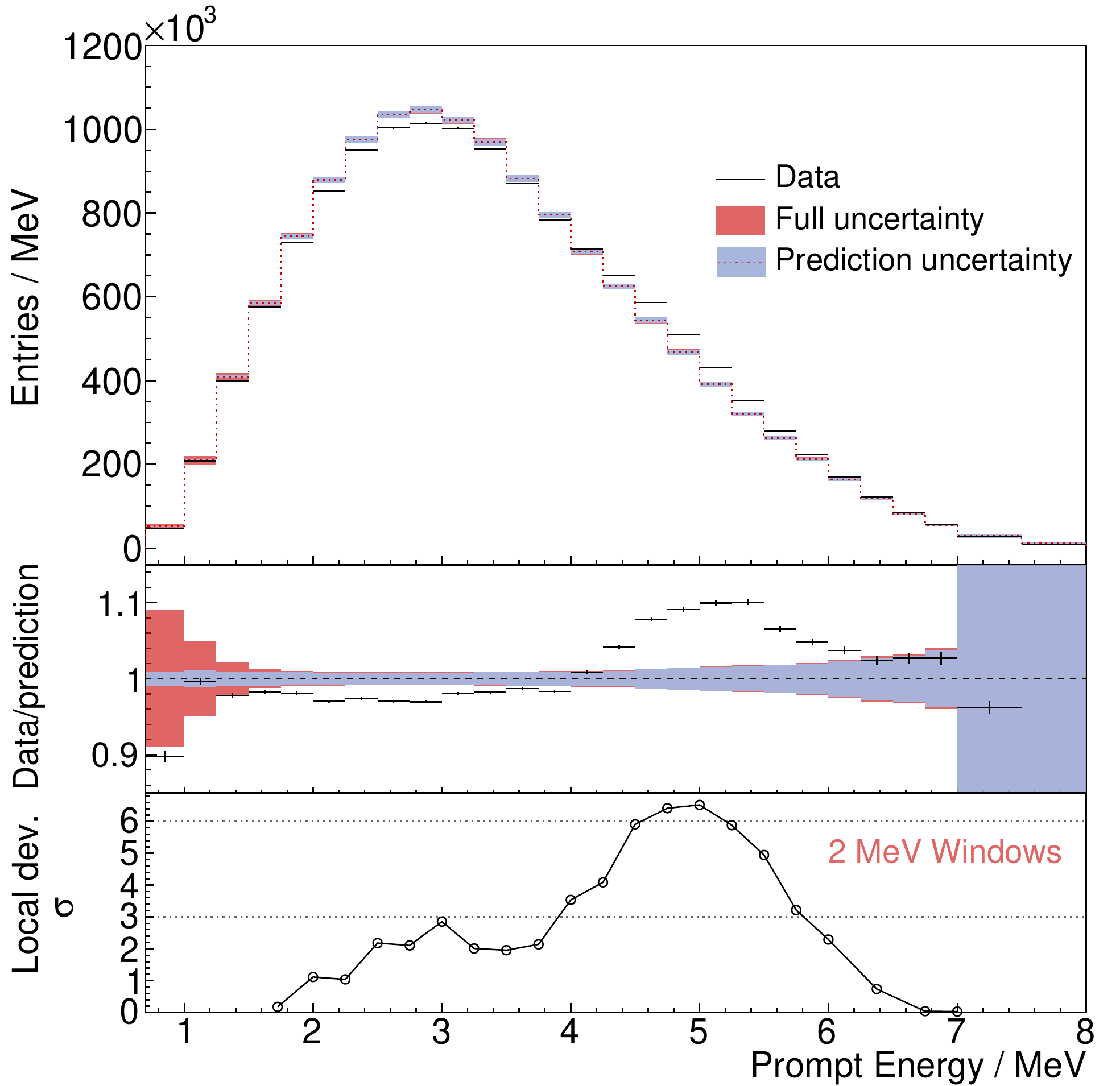}
\caption{
  (Top) Predicted and measured prompt energy spectra.
  The prediction is based on the Huber-Mueller model and is
  normalized to the number of measured events.
  The blue and red filled bands
  represent the square root of diagonal elements of
  the covariance matrix for the flux prediction and
  the full systematic uncertainties, respectively.
  (Middle)
  Ratio of the measured prompt energy spectrum and the normalized predicted spectrum.
  The error bars on the data points represent the statistical uncertainty.
  (Bottom) The local significance of the shape deviation in a sliding 2-MeV window showing a maximum $6.3\sigma$ discrepancy in 4--6~MeV\@.}
\label{total_spectrum}
\end{figure}

Figure~\ref{total_spectrum} shows the spectrum comparison of the measurement with the Huber-Mueller model prediction normalized to the measured number of events.
The measurement and prediction show a large discrepancy particularly near 5~MeV.
With a sliding 2-MeV window scanning following Ref.~\cite{bib:cpc_reactor}, the largest local discrepancy is found in 4--6~MeV, with a significance of 6.3$\sigma$.
The global discrepancy of the entire spectrum in 0.7--8~MeV has a significance of 5.3$\sigma$.

In summary, the IBD yields and prompt energy spectra of $^{235}$U and $^{239}$Pu as the two dominant components in commercial reactors are obtained for the first time using the evolution of the prompt spectrum as a function of fission fractions.
The spectral shape comparison shows similar excesses of events in 4--6~MeV for both $^{235}$U (7\%) and $^{239}$Pu (9\%).
The significance of discrepancy for the $^{235}$U spectral shape is 4.0$\sigma$ while it is 1.2$\sigma$ for the  $^{239}$Pu spectrum due to a larger uncertainty.
In addition, an improved measurement of the prompt energy spectrum of reactor $\bar\nu_e$ is reported with a more precise energy response model and 1958 days of data.
The discrepancy between the measured spectrum shape and the prediction is found to be 5.3$\sigma$ and 6.3$\sigma$ in the entire energy range of 0.7--8~MeV and in a local energy range of 4--6~MeV, respectively.
These discrepancies suggest incorrect spectrum prediction in the Huber-Mueller model, as has been indicated in other theoretical works~\cite{Hayes:2013wra, Fang:2015cma, Sonzogni:2017wxy}.
Direct measurements of the antineutrino flux and spectrum, and the extraction of the $^{235}$U and $^{239}$Pu spectra provide alternative reference spectra for other reactor antineutrino experiments.

This work was supported in part by
the Ministry of Science and Technology of China,
the U.S. Department of Energy,
the Chinese Academy of Sciences,
the CAS Center for Excellence in Particle Physics,
the National Natural Science Foundation of China,
the Guangdong provincial government,
the Shenzhen municipal government,
the China General Nuclear Power Group,
the Research Grants Council of the Hong Kong Special Administrative Region of China,
the Ministry of Education in Taiwan,
the U.S. National Science Foundation,
the Ministry of Education, Youth, and Sports of the Czech Republic,
the Charles University Research Centre UNCE,
the Joint Institute of Nuclear Research in Dubna, Russia,
the National Commission of Scientific and Technological Research of Chile,
We acknowledge Yellow River Engineering Consulting Co., Ltd., and China Railway 15th Bureau Group Co., Ltd., for building the underground laboratory.
We are grateful for the ongoing cooperation from the China Guangdong Nuclear Power Group and China Light~\&~Power Company.

\bibliographystyle{apsrev4-1}
\bibliography{Reactor_Decomposition_PRL}{}
\end{document}